\documentclass[conference]{IEEEtran}
\IEEEoverridecommandlockouts
\usepackage{cite}
\usepackage{amsmath,amssymb,amsfonts}
\usepackage{algorithmic}
\usepackage{graphicx}
\usepackage{textcomp}
\usepackage{xcolor}
\usepackage{subcaption}
\usepackage{booktabs}
\def\BibTeX{{\rm B\kern-.05em{\sc i\kern-.025em b}\kern-.08em
    T\kern-.1667em\lower.7ex\hbox{E}\kern-.125emX}}
\begin{document}

\title{AVENet: Disentangling Features by Approximating Average Features for Voice Conversion \\

\thanks{$^{\ast}$Equal contribution. $^{\dag}$Corresponding author.}
}

\author{\IEEEauthorblockN{Wenyu Wang$^{\ast}$}
\IEEEauthorblockA{\textit{School of Software Engineering} \\
\textit{Xi’an Jiaotong University}\\
Xi'an, China \\
wenyu.wang@stu.xjtu.edu.cn}
\and
\IEEEauthorblockN{Yiquan Zhou$^{\ast}$}
\IEEEauthorblockA{\textit{School of Software Engineering} \\
\textit{Xi’an Jiaotong University}\\
Xi'an, China \\
yiqian.zhou@stu.xjtu.edu.cn}
\and
\IEEEauthorblockN{Jihua Zhu$^{\dag}$}
\IEEEauthorblockA{\textit{School of Software Engineering} \\
\textit{Xi’an Jiaotong University}\\
Xi'an, China \\
zhujh@xjtu.edu.cn}
\and
\IEEEauthorblockN{Hongwu Ding}
\IEEEauthorblockA{\textit{School of Computer Science and Technology} \\
\textit{Anhui University}\\
Hefei, China \\
E22201087@stu.ahu.edu.cn}
\and
\IEEEauthorblockN{Jiacheng Xu}
\IEEEauthorblockA{\textit{School of Software Engineering}\\
\textit{East China Normal University}\\
Shanghai, China \\
xujiacheng28@outlook.com}
\and
\IEEEauthorblockN{Shihao Li}
\IEEEauthorblockA{\textit{Division of Music and Audio} \\
\textit{Union Wheatland Culture and Media Ltd.}\\
Chengdu, China \\
yiseho@yiseho.com}
}
\maketitle

\begin{abstract}
    Voice conversion (VC) has made progress in feature disentanglement, but it is still difficult to balance timbre and content information. This paper evaluates the pre-trained model features commonly used in voice conversion,  and proposes an innovative method for disentangling speech feature representations. Specifically, we first propose an ideal content feature, referred to as the average feature, which is calculated by averaging the features within frame-level aligned parallel speech (FAPS) data. For generating FAPS data, we utilize a technique that involves freezing the duration predictor in a Text-to-Speech system and manipulating speaker embedding. To fit the average feature on traditional VC datasets, we then design the AVENet to take features as input and generate closely matching average features. Experiments are conducted on the performance of AVENet-extracted features within a VC system. The experimental results demonstrate its superiority over multiple current speech feature disentangling methods. These findings affirm the effectiveness of our disentanglement approach.
\end{abstract}


\section{Introduction}
\label{sec:intro}

Voice conversion (VC) is a technology designed to transform source speech into target speech while preserving the original content and expressiveness\cite{mohammadi2017overview, liu2020transferring}. With advancements in deep learning, voice conversion has recently achieved significant progress\cite{wang2023delivering,lian2022towards, polyak2021speech}. A key challenge in VC is the effective separation and recombination of speaker timbre from content information. This extraction process has a substantial impact on the overall performance of VC systems.

Researchers have proposed a variety of approaches to achieve the disentanglement of content information from timbre information. Some have sought to adapt methods from image style transfer tasks, utilizing Generative Adversarial Networks (GANs)\cite{kaneko2018cyclegan, li2021starganv2} for voice conversion tasks. However, due to the intrinsic instability of GANs, these systems frequently generate speech that lacks clarity. Other researchers have explored the use of vector quantization\cite{huang2020unsupervised}, adaptive instance normalization\cite{chen2021again} and information perturbation\cite{xie2022end} to disentangle content information from speech. Despite these efforts, the disentanglement methods often remain imperfect. This can result in a decrease in the naturalness and speaker similarity of the converted speech.

Recently, researchers have utilized data-pretrained models to extract robust linguistic representations that encompass content information. These models have been widely adopted in various downstream tasks that emphasize content understanding. Currently, the application of pre-trained models in VC systems can be primarily categorized into two types: those based on Automatic Speech Recognition (ASR) models and those based on self-supervised learning (SSL) models.

VC methods based on ASR models\cite{yao2021wenet, radford2023robust} utilize intermediate content representations derived from ASR systems, such as phoneme posteriorgrams (PPGs) and bottleneck features (BNFs) \cite{sun2016phonetic, zhao2022disentangling, wang2021enriching}. The primary goal of ASR tasks is to extract semantic information from speech. As a result, the training methods effectively isolate content information. However, these methods necessitate substantial amounts of labeled data to train the ASR model. Furthermore, the accuracy and granularity of data labeling influence the model's performance.

\begin{figure*}[t]
  \centering
  \begin{subfigure}[b]{.36\linewidth}
    \centering
    \includegraphics[width=\textwidth]{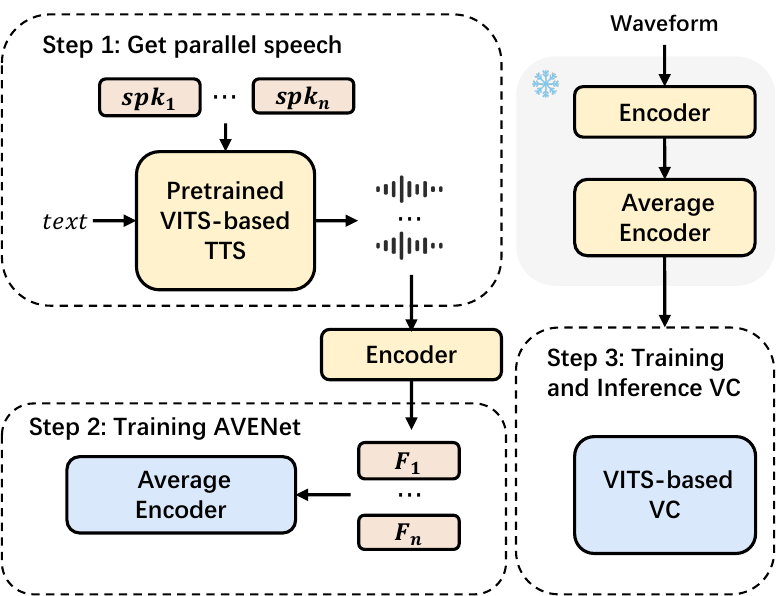}
    \caption{Overall process}
    \label{fig:1(a)}
  \end{subfigure}
  \hfill
  \begin{subfigure}[b]{.34\linewidth}
    \centering
    \includegraphics[width=\textwidth]{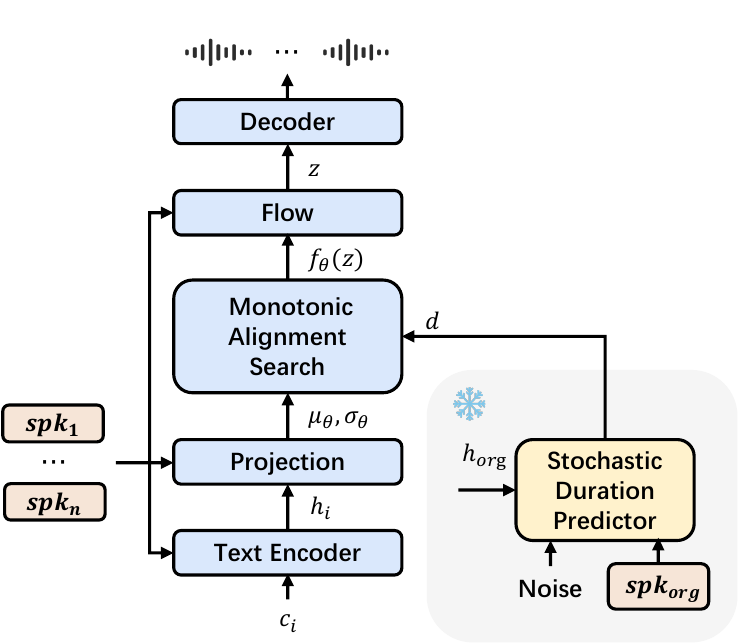}
    \caption{Parallel speech inference}
    \label{fig:1(b)}
  \end{subfigure}
  \hfill
  \begin{subfigure}[b]{.20\linewidth}
    \centering
    \includegraphics[width=\textwidth]{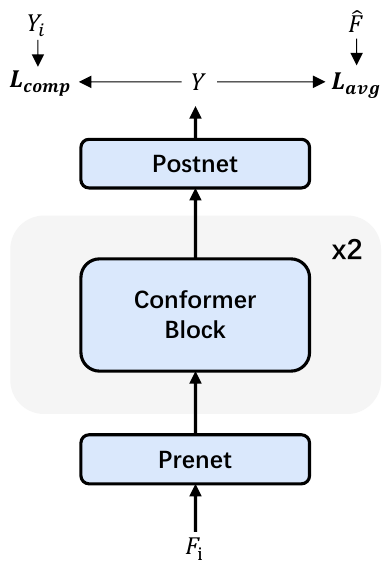}
    \caption{AVENet}
    \label{fig:1(c)}
  \end{subfigure}
  \caption{The overall architecture of average features for disentangling speech information includes three main components: (a) Overall process, (b) FAPS inference, and (c) AVENet.}
  \label{fig:1}
\end{figure*}

Due to the superior performance of SSL models\cite{baevski2020wav2vec, hsu2021hubert, chen2022wavlm} in downstream ASR tasks, it has been convincingly demonstrated that these models can extract robust content information from speech data. Consequently, the second category of research focuses on extracting content information from features derived from SSL models. FreeVC\cite{li2023freevc} employs a spectrogram-resize (SR) operation to distorts speaker information without altering content information, thereby enhancing the model's disentanglement capability. The adversarial speaker disentanglement technique\cite{zhao2023adversarial} reduces residual speaker information in SSL features. This is achieved through adversarial training methods and the use of an external unlabeled corpus. SoftVC\cite{van2022comparison} and ContentVec\cite{qian2022contentvec} further disentangle content information from HuBERT features by introducing Hubert-soft units and an SSL teacher-student framework, respectively. However, the first two approaches lack sufficient separation between speaker timbre and content information, limiting the VC system's ability to accurately convert the timbre. Furthermore, these latter three methods may inadvertently eliminate content information while attempting to remove speaker timbre. This can affect the naturalness of the VC system's output. These limitations hinder the performance of voice conversion. A more effective disentanglement technique is needed to retain ample content information in speech while minimizing speaker timbre information as much as possible.

Today, Text-to-Speech (TTS) technology, which converts text into speech in the voice of a specific speaker, has advanced to the point where they can produce high-quality speech that rivals human pronunciation. This development makes it feasible to use TTS technology as a tool for data augmentation.

In this paper, we propose an ideal content feature referred to as the average feature. Previous research has discussed the concept of average mel-spectrograms\cite{popov2021diffusion, zhu2024noise}, but these features are limited to the phoneme level, resulting in significant information loss. In contrast, our goal is to obtain a finer-grained, frame-level average feature. The average features are obtained through the following method: First, features are extracted from frame-aligned parallel speeches (FAPS), where the frames at the same time points have identical content information but different timbres. Then, these features are averaged. This process can be based on intermediate content features from ASR models or SSL features. However, in reality, it is very difficult to find FAPS. To obtain FAPS data, we utilize TTS system capable of synthesizing highly natural speech \cite{kim2021conditional}, generating FAPS data by freezing the duration predictor while altering the speaker embeddings. To use average features in scenarios like VC where FAPS data is not available, we have developed the AVENet module to fit the average features. During the training phase, AVENet leverages the FAPS data to learn how to transform raw speech features into outputs that are highly similar to the average features. We introduced a reconstruction loss and a positive contrastive loss to achieve this. In the inference or application phase, AVENet can be directly applied to traditional VC datasets. It is capable of extracting representations similar to the average features from the speech of a single speaker, even if these speech samples are not from FAPS. 

In the experiments, we evaluated the timbre information leakage of three widely used pre-trained model features: WavLM\cite{chen2022wavlm}, HuBERT\cite{hsu2021hubert}, and Whisper\cite{radford2023robust}. We trained AVENet using features extracted from each of these pre-trained models to assess its disentanglement capabilities. Building on this foundation, we conducted comprehensive VC experiments utilizing the disentangled features. The experimental results indicate that AVENet effectively approximates the average feature. It successfully eliminates speaker timbre information while preserving the original content and expressiveness.

\section{Proposed Methods}

In this section, we will provide a comprehensive exposition of our disentanglement method. We introduce an ideal content feature referred to as the average feature in subsection A. Fig.\ref{fig:1(a)} illustrates the process of utilizing this feature for disentanglement and its application within the VC system. In subsection B, we offer an in-depth analysis of the first step illustrated in the figure, which involves obtaining frame-aligned parallel speeches (FAPS) data. The second step is elaborated upon in subsection C, where we outline the architecture of the AVENet and its training process to translate raw features into our defined average features. Finally, in subsection D, we describe the third step: a VITS-based VC system designed to compare and analyze the performance of various features.

\subsection{Average feature}
Consider a feature sequence $F = [F_1; \ldots; F_t]$ representing speech, where $F_t$ is the speech feature vector for frame $t$. For a group of $N$ aligned sets of FAPS features, denoted by $\{F^i\}_{i=1}^{N}$, where each $F^i = [F_1^i;\ldots;F_t^i]$ represents the features of the $i$-th speech. Each $F^i$ contains distinct timbre information while maintaining consistent content information. When the number of $N$ is sufficiently large, we define the mean of each set $\{F^i\}_{i=1}^{N}$ as the average feature $\hat{F}$, which serves as an ideal content feature. $\hat{F}$ is expressed as:
\begin{align}
  \hat{F}=\frac{1}{N} \sum_{i=1}^N F^{i}=\frac{1}{N} \sum_{i=1}^N\left[F_{1}^i ; \ldots ; F_{t}^i\right]
\end{align}

\subsection{Synthesis frame-level aligned parallel speech}

As mentioned earlier, the FAPS refers to speech data where the content information is identical at the frame level, while different speaker timbres are preserved. Due to variations in rhythm and pause durations caused by individual pronunciation habits, obtaining this information through manual recording is impractical.

To tackle this, we utilized VITS\cite{kim2021conditional,kong2023vits2} to generate FAPS data. VITS proposes a parallel, end-to-end TTS technique that performs comparably to human speech in Mean Opinion Score evaluations. In multi-speaker speech synthesis, each speaker is associated with distinct speaker embeddings ($spk$), which are integrated into the VITS modules to generate speech with unique speaker characteristics. VITS employs a stochastic duration predictor to estimate the duration of phonemes. This predictor is conditioned on inputs such as the text representation ($h_{text}$) and $spk$. During the synthesis process, VITS can generate speech with natural variations in duration.

We modified the inference process of the VITS system. We randomly select a speaker from the pool of available speakers, denoted as $org$. The speaker embedding $spk_{org}$ is combined with the text representation $h_{org}$ to generate the phoneme duration $d$. Inference is performed for each speaker's $spk$ along with $d$ and $h_{org}$ to generate ${X^i}$, where $i$ represents the $i$-th speaker, and ${X^i}$ represents the $i$-th speech in the FAPS. Fig.\ref{fig:1(b)} illustrates this inference process.

To promote speaker diversity, we employ a linear weighting strategy (LWS). This strategy blends multiple speaker embeddings during inference to generate speech with varied speaker characteristics. In Equation \ref{eq:spkw}, $w^i$ represents the weight for the $i$-th speaker, and the weight constraint $\sum_iw^i = 1$ is enforced.
\begin{align}
  spk^w = \sum_iw^i\cdot spk^i
  \label{eq:spkw}
\end{align}

\subsection{AVENet}
In voice conversion tasks, the training and inference processes utilize real speech data. Directly obtaining the desired average feature by simply averaging the data, as done in FAPS, is not feasible in this context. Therefore, we introduce a network structure denoted as $f$, called AVENet. This network approximates the features of actual speech with average features, thereby disentangling the content information.

During the training phase of AVENet, we utilized the synthetic FAPS data described in the previous section. Let $F$ represent the raw features as input. Our objective is to ensure that $Y=f(F)$ closely approximates the average features $\hat{F}$, while the outputs $Y_i$ of FAPS should be more similar. As illustrated in Fig.\ref{fig:1(c)}, AVENet consists of two conformer blocks\cite{gulati2020conformer} and a residual network architecture.

To fit $\hat{F}$, we introduce the average reconstruction loss ($\mathcal{L}_{avg}$). This loss quantifies the difference between $Y$ and $\hat{F}$, defined as the $L_1$ loss between the two. It encourages the encoder to produce features that closely approximate the desired average features. To further promote the removal of speaker timbre information from the output features of AVENet, we introduce a positive contrastive loss. During training, we randomly select different features $F^i$ from the FAPS set $\{F^i\}_{i=1}^{N}$ and obtain the corresponding output $Y^i$ through AVENet. The positive contrastive loss $\mathcal{L}_{comp}$ is then defined as the $L_1$ loss between $Y$ and $Y^i$. This loss encourages the output features of FAPS after passing through AVENet to be more similar. The total loss is represented as follows:
\begin{align}
  \mathcal{L}= \alpha\mathcal{L}_{avg} + \beta \mathcal{L}_{comp}
\end{align}

Where $\alpha$ and $\beta$ are weight coefficients. Through extensive experiments, we found that setting their values to 1.0 and 0.5, respectively, yields better training results.

\subsection{VITS-based VC}

\begin{figure}[htb]
  \centering
  \centerline{\includegraphics[width=8.5cm]{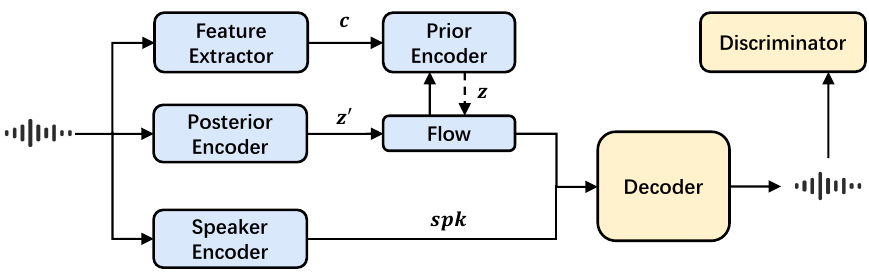}}
\caption{Architecture of VITS-based VC.}
\label{fig:vits}
\end{figure}

Our VITS-based VC framework integrates a posterior encoder, prior encoder, decoder, discriminator, and speaker encoder. Apart from the feature extractor module, the VC architecture is identical to FreeVC\cite{li2023freevc}. Fig.\ref{fig:vits} illustrates the architecture of our VC system.

\section{Experiments}

\begin{table*}
  \caption{The subjective evaluation results are presented in terms of MOS with 95\% confidence intervals, corresponding to the scenarios of "seen-to-seen" and "unseen-to-unseen". The objective evaluation results included CER and Cos.Sim.}
  \label{tab:1}
  \begin{tabular}
   {c@{\hspace{20pt}}|c@{\hspace{20pt}}|c@{\hspace{22pt}}c@{\hspace{22pt}}|c@{\hspace{22pt}}c@{\hspace{22pt}}|c@{\hspace{20pt}}c@{\hspace{20pt}}}
    \hline &  & \multicolumn{2}{c|}{ seen-to-seen } & \multicolumn{2}{c|}{ unseen-to-unseen } & \multicolumn{2}{c}{ objective } \\
    \midrule
    Features & Approach & Naturalness $\uparrow$ & Similarity $\uparrow$ & Naturalness $\uparrow$ & Similarity $\uparrow$ & CER $\downarrow$ & Cos.Sim $\uparrow$ \\
    \midrule
     & Origin & $3.82 \pm 0.08$ & $3.34 \pm 0.08$ & $\mathbf{3 . 9 0} \pm \mathbf{0 . 0 5}$ & $3.06 \pm 0.09$ & $8.65\%$ & $0.8303$ \\
     & SR & $3.80 \pm 0.07$ & $3.69 \pm 0.07$ & $3.85 \pm 0.06$ & $3.47 \pm 0.09$ & $8.01\%$ & $0.8397$ \\
     Whisper & GANs & $3.73 \pm 0.07$ & $3.45 \pm 0.08$ & $3.78 \pm 0.07$ & $3.25 \pm 0.07$ & $7.95\%$ & $0.8308$ \\
     & Perturb & $3.61 \pm 0.08$ & $3.86 \pm 0.07$ & $3.74 \pm 0.06$ & $3.57 \pm 0.06$ & $9.43\%$ & $0.8423$ \\
     & AVENet & $\mathbf{3 . 8 7} \pm \mathbf{0 . 09}$ & $\mathbf{4 . 1 8} \pm \mathbf{0 . 0 5}$ & $3.82 \pm 0.06$ & $\mathbf{3 . 8 1} \pm \mathbf{0 . 0 8}$ & $\mathbf{7.06\%}$ & $\mathbf{0.8488}$ \\
    \midrule
     & Origin & $3.75 \pm 0.09$ & $3.36 \pm 0.07$ & $3.76 \pm 0.07$ & $3.11 \pm 0.08$ & $13.89\%$ & $0.8341$ \\
     & SR & $3.79 \pm 0.08$ & $3.71 \pm 0.07$ & $3.75 \pm 0.07$ & $3.30 \pm 0.07$ & $10.91\%$ & $0.8414$ \\
     & GANs & $3.48 \pm 0.08$ & $3.56 \pm 0.08$ & $3.58 \pm 0.08$ & $3.20 \pm 0.07$ & $11.05\%$ & $0.8433$ \\
     HuBERT & Perturb & $3.53 \pm 0.07$ & $3.84 \pm 0.08$ & $3.49 \pm 0.08$ & $3.49 \pm 0.06$ & $12.64\%$ & $0.8476$ \\
     & Hu-soft & $\mathbf{3 . 9 6} \pm \mathbf{0 . 0 7}$ & $3.47 \pm 0.07$ & $3.76 \pm 0.08$ & $3.28 \pm 0.07$ & $10.38\%$ & $0.8391$ \\
     & ContentVec & $3.86 \pm 0.08$ & $4.02 \pm 0.08$ & $3.70 \pm 0.08$ & $3.70 \pm 0.07$ & $\mathbf{9.09\%}$ & $0.8443$ \\
     & AVENet & $3.91 \pm 0.09$ & $\mathbf{4 . 2 4} \pm \mathbf{0 . 0 7}$ & $\mathbf{3 . 7 7} \pm \mathbf{0 . 0 7}$ & $\mathbf{3 . 9 1} \pm \mathbf{0 . 0 9}$ & $9.95\%$ & $\mathbf{0.8521}$ \\
    \midrule
     & Origin & $4.03 \pm 0.09$ & $2.80 \pm 0.10$ & $3.87 \pm 0.06$ & $2.42 \pm 0.09$ & $9.98\%$ & $0.7858$ \\
    & SR & $4.06 \pm 0.09$ & $3.20 \pm 0.09$ & $3.85 \pm 0.07$ & $2.75 \pm 0.10$ & $8.78\%$ & $0.8205$ \\
     WavLM & GANs & $3.77 \pm 0.09$ & $3.00 \pm 0.10$ & $3.63 \pm 0.07$ & $2.64 \pm 0.10$ & $9.76\%$ & $0.7955$ \\
     & Perturb & $3.65 \pm 0.08$ & $3.74 \pm 0.07$ & $3.59 \pm 0.09$ & $3.54 \pm 0.08$ & $10.65\%$ & $0.8319$ \\
     & AVENet & $4.22 \pm 0.09$ & $\mathbf{4 . 2 8} \pm \mathbf{0 . 0 7}$ & $\mathbf{3 . 9 5} \pm \mathbf{0 . 0 7}$ & $\mathbf{3 . 8 7} \pm \mathbf{0 . 0 9}$ & $6.52\%$ & $\mathbf{0.8382}$ \\
     & - $\mathcal{L}_{comp}$ & $\mathbf{4.23} \pm \mathbf{0.06}$ & $3.97 \pm 0.09$ & $3.75 \pm 0.07$ & $3.59 \pm 0.08$ & $\mathbf{6.33\%}$ & $0.8327$ \\
     & - LWS & $4.11 \pm 0.08$ & $4.18 \pm 0.06$ & $3.74 \pm 0.08$ & $3.51 \pm 0.07$ & $8.08\%$ & $0.8315$ \\

    \midrule
  \end{tabular}
\end{table*}

\subsection{Experimental setup}
All experiments were conducted using the AISHELL3\cite{shi2021aishell} dataset, a large-scale, high-fidelity multilingual Mandarin corpus that includes data from 218 Mandarin speakers. When partitioning the dataset, we allocated all data from 10 speakers entirely to the test set to evaluate the method's capability in zero-shot scenarios. From the remaining 208 speakers, we randomly selected 2 samples from each speaker for the validation and 10 samples for the test set, with the remainder being used as the training set.

The FAPS data utilized in the experiments was synthesized using the BERT-VITS2 system. We trained a BERT-VITS2\footnote{https://github.com/fishaudio/Bert-VITS2} model on the previously mentioned training set and fine-tuned it for specific 20 speakers. We synthesized a total of 15,000 FAPS sets, with each set containing 40 speakers: 20 speakers obtained directly and an additional 20 speakers acquired through LWS. Notably, the average features were computed using data from only 20 speakers obtained directly, as their speaker timbre information for each FAPS was fixed.

All audio samples are downsampled to 16 kHz. The FFT, window, and hop size are set to 1280, 1280, and 320, respectively. We conducted experiments on 1024-dimensional WavLM, 256-dimensional HuBERT, and 1280-dimensional Whisper features, respectively. For each type of feature, the corresponding AVENet model was trained on the FAPS data generated previously for 300,000 steps on an A100 GPU with a batch size of 64. The features for the VITS-based VC system were derived from the AVENet model and were trained using the training set. All VC experiments utilized the same VC system architecture, were trained for 500,000 steps on an A100 GPU with a batch size of 64. We highly recommend that readers listen to our samples\footnote{https://ryker-icme2025.github.io/avenet/}.

\subsection{FAPS data quality}

We designed a metric $E_{avg}$ to demonstrate the alignment quality of the synthetically generated FAPS data. We also evaluated the real data and the synthesized FAPS using Mean Opinion Score (MOS).

We randomly selected 1000 pairs of speech from the FAPS data. We used the Montreal Forced Aligner (MFA)\cite{mcauliffe2017montreal} to extract the timestamps of the phonemes. Let $n$ denote the number of parallel speech pairs, $P_i$ be the $i$-th pair of parallel speech, and $S_i$ be the set of phonemes in $P_i$. $e_{start}(s)$ and $e_{end}(s)$ represent the starting and ending errors for each phoneme $s \in S_i$. We calculated the average start and end errors for all phonemes in the parallel speech as the evaluation metric:
\begin{align}
E_{avg} = \frac{1}{n} \sum_{i=1}^{n}\frac{1}{|S_i|} \sum_{s \in S_i} \frac{e_{start}(s) + e_{end}(s)}{2}
\end{align}
The computed average error $E_{avg}$ for the phonemes in FAPS data was only \textbf{0.011s}, indicating excellent alignment performance.

\begin{table}[ht]
    \caption{The MOS evaluation for FAPS and real speeches.}
    \label{table:compare_baseline}
    \centering
    \renewcommand{\arraystretch}{1.2}
    \begin{tabular}{l@{\hspace{27pt}}|c@{\hspace{26pt}}|c@{\hspace{26pt}}}
        \hline
        Classification & Naturalness & Similarity \\
        \hline
        Ground Truth Dataset& $4.31 \pm 0.05$ & $4.48 \pm 0.06$ \\
        FAPS Dataset& $3.95 \pm 0.07$ & $4.42 \pm 0.05$ \\
        \hline
    \end{tabular}
\end{table}

The detailed rules for the MOS evaluation are consistent with those described in the VC experimental results section. The results are presented in Table \ref{table:compare_baseline}. The generated data exhibits excellent naturalness and an extremely high degree of speaker similarity, rivaling that of real speech. This indicates the high quality of the FAPS dataset.

\subsection{Disentangle effectiveness of AVENet}

We extracted 1,000 pairs of original FAPS features and their corresponding features processed through AVENet. We calculated the Mean Absolute Value (MAV) for each type of feature. To demonstrate AVENet's capability in disentangling speaker timbre information, we also measured the distance between each pair of original FAPS features and their corresponding features after processing by AVENet. The results are presented in Table \ref{table:distance}. After processing with AVENet, the distance between FAPS features for each type was reduced by more than three times, indicating that AVENet enhances the similarity of features with the same content. Additionally, we found that among the models, Whisper exhibited the smallest ratio of feature distance to the MAV of the original features, followed by HuBERT, while WavLM had the largest ratio. This indirectly suggests that Whisper features contain the least amount of timbre information. In contrast, WavLM features contain the most, leading to the most significant timbre leakage when applied to voice conversion.

\begin{table}[ht]
    \caption{The average distance between each pair of FAPS features (before and after passing through AVENet)}
    \label{table:distance}
    \centering
    \renewcommand{\arraystretch}{1.2}
    \begin{tabular}{l@{\hspace{20pt}}|c@{\hspace{15pt}}|c@{\hspace{15pt}}|c@{\hspace{15pt}}}
        \hline
        Features & MAV & distence (origin) & distence (AVENet)\\
        \hline
        Whisper & $0.629$ & $0.194$ & $0.078$\\
        HuBERT & $0.693$ & $0.301$ & $0.109$\\
        WavLM & $0.152$ & $0.092$ & $0.015$\\
        \hline
    \end{tabular}
\end{table}

\begin{figure}[htb]
  \centering
  \begin{subfigure}[b]{.48\linewidth}
    \centering
    \includegraphics[width=0.9\textwidth]{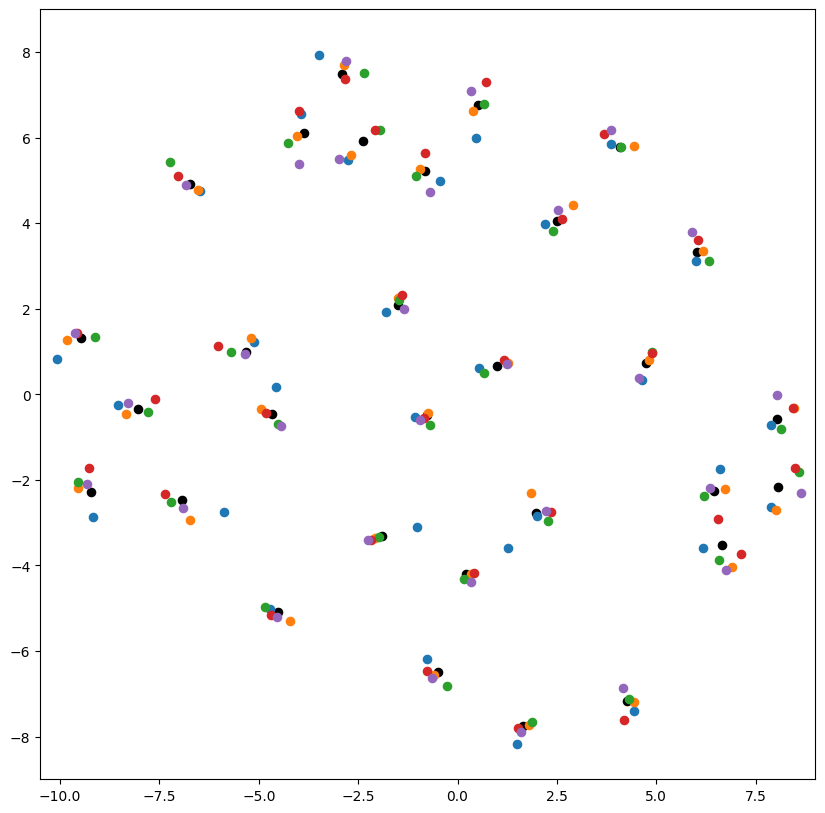}
    \caption{t-SNE with WavLM}
    \label{fig:2a}
  \end{subfigure}
  \begin{subfigure}[b]{.48\linewidth}
    \centering
    \includegraphics[width=0.9\textwidth]{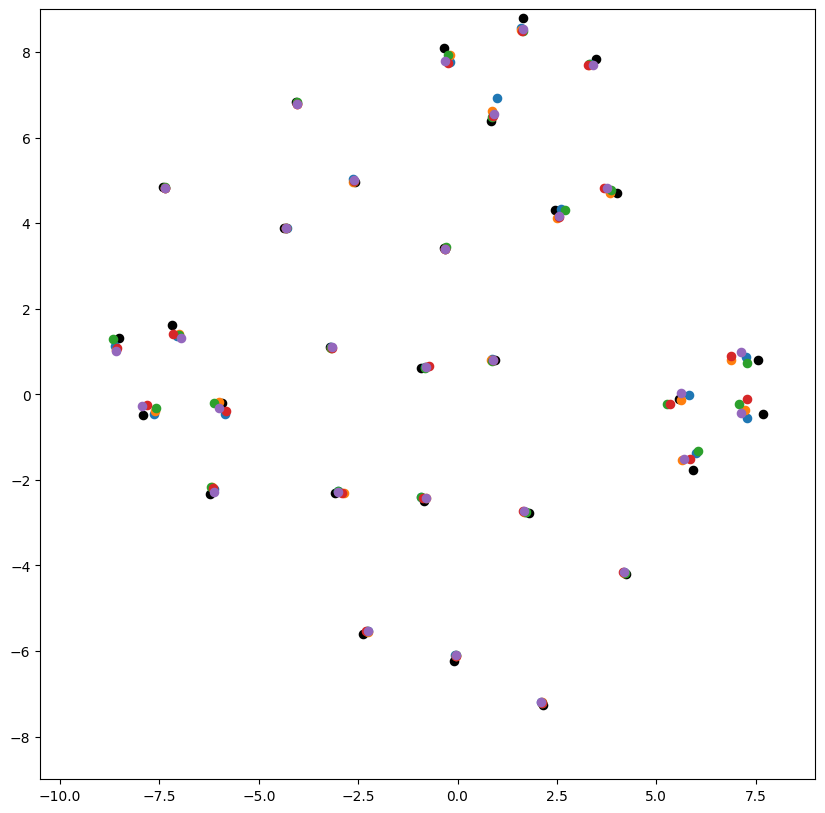}
    \caption{ WavLM-AVENet}
    \label{fig:2b}
  \end{subfigure}
  \begin{subfigure}[b]{.48\linewidth}
    \centering
    \includegraphics[width=0.9\textwidth]{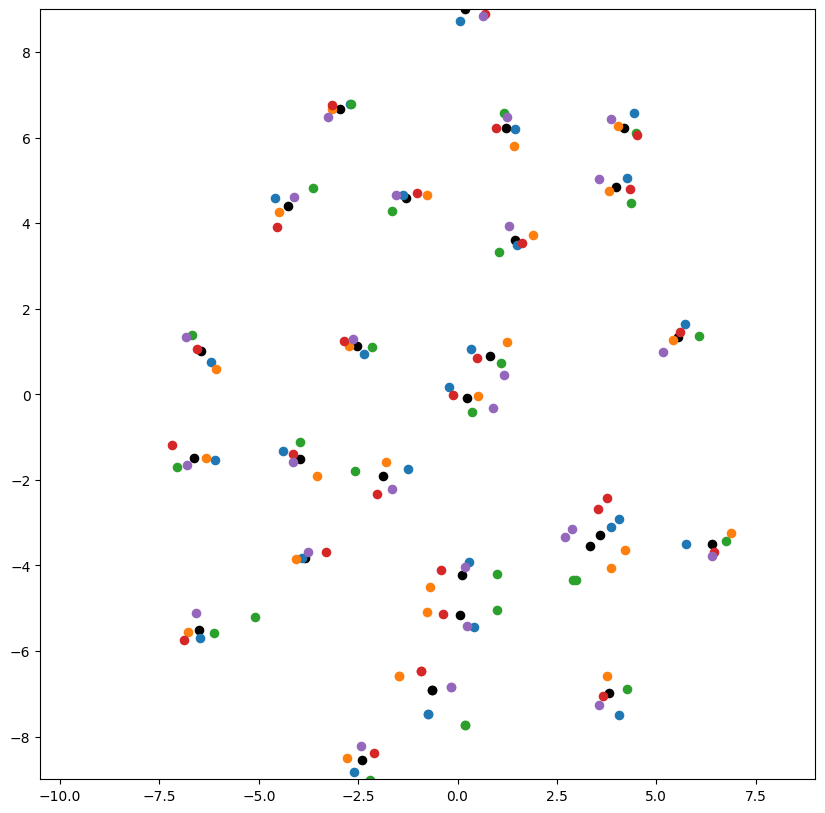}
    \caption{t-SNE with HuBERT}
    \label{fig:2a}
  \end{subfigure}
  \begin{subfigure}[b]{.48\linewidth}
    \centering
    \includegraphics[width=0.9\textwidth]{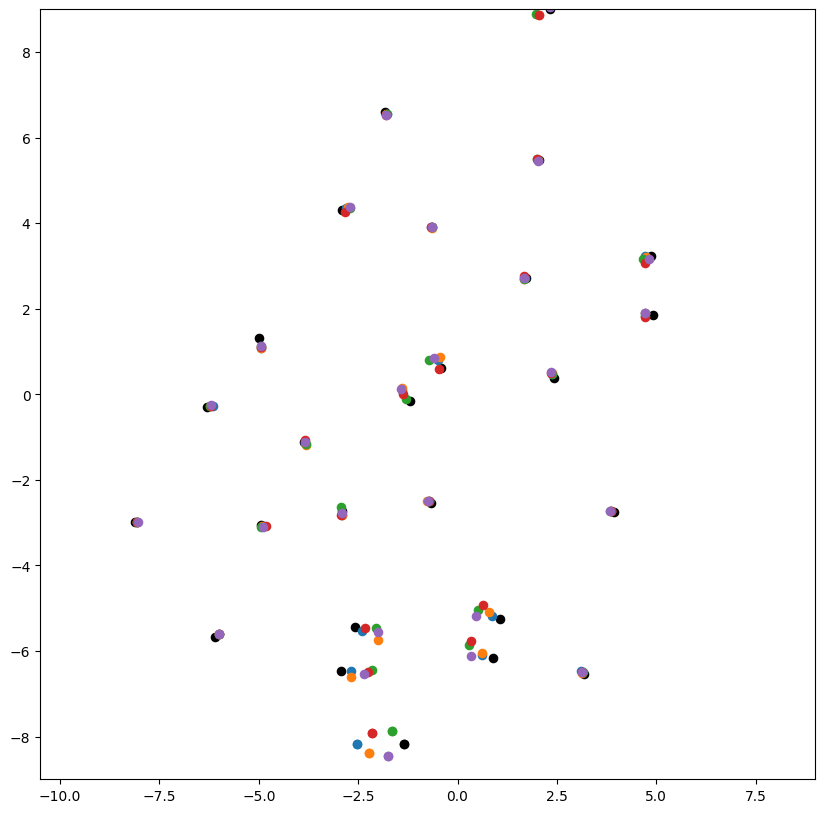}
    \caption{HuBERT-AVENet}
    \label{fig:2b}
  \end{subfigure}
  \begin{subfigure}[b]{.48\linewidth}
    \centering
    \includegraphics[width=0.9\textwidth]{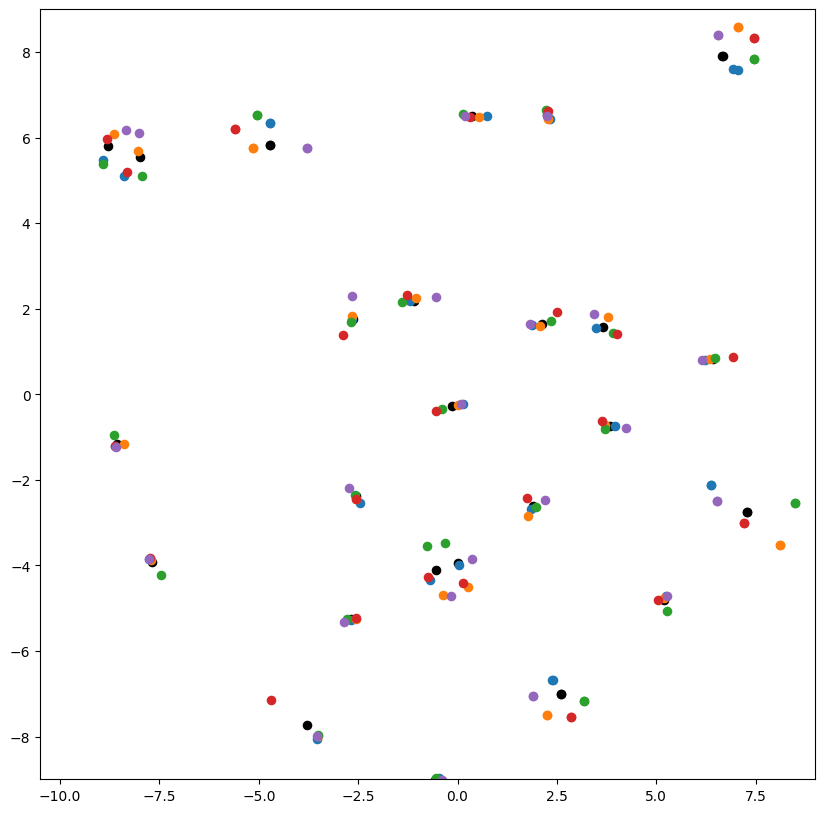}
    \caption{t-SNE with Whisper}
    \label{fig:2a}
  \end{subfigure}
  \begin{subfigure}[b]{.48\linewidth}
    \centering
    \includegraphics[width=0.9\textwidth]{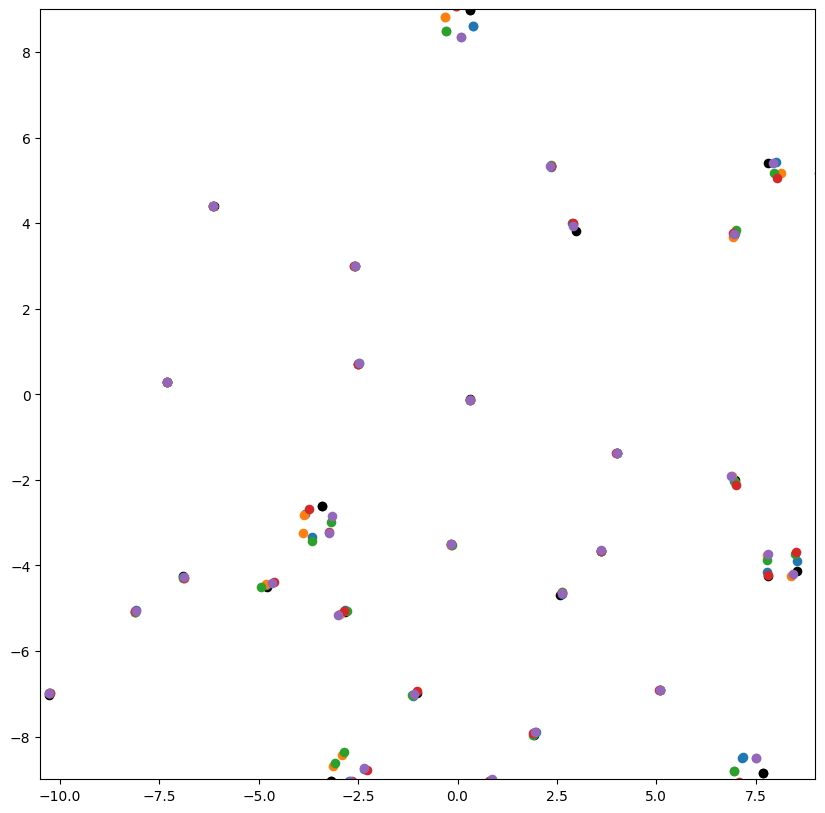}
    \caption{Whisper-AVENet}
    \label{fig:2b}
  \end{subfigure}
  \caption{Visualizing the output of AVENet using t-SNE.}
  \label{fig:2}
\end{figure}

We visualized three types of features on the graph. 
We selected the original features, AVENet output features, and the average features for 30 frames from each of five source speakers. We used t-SNE\cite{van2008visualizing} to project these features into two dimensions. Fig.\ref{fig:2} illustrates the projection of the features. The black dots represent the average features, while each color corresponds to a different speaker. After applying AVENet, the features from the same frame are more closely clustered in the graph and are also nearer to the average feature.

\subsection{VC Experimental Results}

We compared the disentanglement methods for three types of features processed by AVENet with the original features. We also included several disentanglement techniques in the comparison: the SR operation\cite{li2023freevc}, GANs operation, and information perturbation operation. The GANs operation uses a discriminator to distinguish the speaker information from the features. The information perturbation method is based on the approach mentioned in NANSY\cite{choi2021neural} for feature extraction. We also compared disentangled HuBERT features with HuBERT-soft\cite{van2022comparison} and ContentVec\cite{qian2022contentvec}. We evaluated the performance of these methods from both subjective and objective perspectives. For subjective evaluation, we invited 20 participants to rate the naturalness and similarity of the speech on a 5-point MOS scale. The MOS scores covered evaluation scenarios including "seen-seen" and "unseen-unseen". For objective evaluation, we used CER and Cosine Similarity (Cos.Sim). CER is measured using an ASR model\footnote{https://github.com/wenet-e2e/wenet} to evaluate the character error rate between the source speech and the converted speech. Cos.Sim is the cosine similarity between the speaker embeddings extracted from the converted speech and the target speaker embeddings. We used the speaker embedding mentioned in PPGVC\cite{Liu2021} to quantify the speaker similarity between two speeches.

Table \ref{tab:1} shows the experimental results. Before discussing the disentanglement effects, we note that using the original features without disentanglement in VC does not achieve the best naturalness, likely due to the entangled information hindering effective model learning.

In subjective evaluations, our proposed method demonstrated commendable speech naturalness and achieved the highest speaker similarity across all three features. Methods based on the SR operation, GANs operation, and information perturbation failed to completely eliminate speaker information. As a result, these methods led to lower speaker similarity. Moreover, methods relying on GANs and information perturbation experienced some information loss, which diminished the naturalness of the speech. In contrast, our method significantly improved speaker similarity while maintaining a high level of speech naturalness. In the "unseen-to-unseen" experiment, our method also exhibited excellent performance, underscoring its strong robustness. In objective evaluations, our method effectively retained the semantic content of the source speech, performing exceptionally well in CER tests. Furthermore, our method achieved a high Cos.Sim, indicating its capability to effectively remove speaker timbre information. We also conducted ablation experiments. The results indicate that incorporating the positive contrastive loss $\mathcal{L}_{comp}$ further reduced the distance between features with similar content. This enhancement led to improved speaker similarity. The LWS method successfully increased the robustness of AVENet, improving both the naturalness and speaker similarity.

\section{Conclusion}

This paper proposes a method for achieving feature disentanglement through the fitting of average features. We evaluated the retention of timbre information in commonly used pre-trained model features, and proposed a specific process for timbre disentanglement utilizing an averaging method. To achieve this, we introduced the FAPS synthesis method and developed and trained AVENet to remove timbre information from the features. Experimental results demonstrated the advantages of our approach. In future research, we will continue to investigate the disentanglement of content information.

\bibliographystyle{IEEEbib}
\bibliography{icme2025references}

\end{document}